# Robust Group Comparison Using Non-Parametric Block-Based Statistics

**Geng Chen** [1], **Pei Zhang** [1], **Ke Li** [2], **Chong-Yaw Wee** [1], **Wenliang Pan** [3], **Yafeng Wu** [4], **Panteleimon Giannakopoulos** [5,6], **Sven Haller** [7,8], **Dinggang Shen** [1,9,*] and **Pew-Thian Yap** [1,*]

[1] *Department of Radiology and Biomedical Research Imaging Center (BRIC), University of North Carolina at Chapel Hill, NC, U.S.A.*
[2] *Fundamental Science on Ergonomics and Environment Control Laboratory, Beihang University, Beijing, China*
[3] *Department of Statistical Science, School of Mathematics, Sun Yat-Sen University, Guangzhou, China*
[4] *Data Processing Center, Northwestern Polytechnical University, Xi'an, China*
[5] *Department of Mental Health and Psychiatry, University Hospitals of Geneva, Switzerland*
[6] *Department of Psychiatry, University of Geneva, Switzerland*
[7] *Affidea CDRC - Centre Diagnostique Radiologique de Carouge, Switzerland*
[8] *Department of Surgical Sciences, Radiology, Uppsala University, Uppsala, Sweden*
[7] *Department of Brain and Cognitive Engineering, Korea University, Seoul, Korea*

Correspondence*:
Pew-Thian Yap and Dinggang Shen
ptyap@med.unc.edu and dgshen@med.unc.edu

## ABSTRACT

Voxel-based analysis methods localize brain structural differences by performing voxel-wise statistical comparisons on two groups of images aligned to a common space. This procedure requires highly accurate registration as well as a sufficiently large dataset. However, in practice, the registration algorithms are not perfect due to noise, artifacts, and complex structural variations. The sample size is also limited due to low disease prevalence, recruitment difficulties, and demographic matching issues. To address these issues, in this paper, we propose a method, called block-based statistic (BBS), for robust group comparison. BBS consists of two major components: Block matching and permutation test. Specifically, based on two group of images aligned to a common space, we first perform block matching so that structural misalignments can be corrected. Then, based on results given by block matching, we conduct robust non-parametric statistical inference based on permutation test. Extensive experiments were performed on synthetic data and the real diffusion MR data of mild cognitive impairment patients. The experimental results indicate that BBS significantly improves statistical power, notwithstanding the small sample size.

## 1 INTRODUCTION

Voxel-based analysis (VBA) methods (Ashburner and Friston, 2000, 2001; Davatzikos et al., 2001) detect group differences in two sets of MR images that have been registered to a common space by performing voxel-wise statistical comparisons. Compared with methods based on regions of interest (ROIs), VBA does not require explicit delineation of ROIs and hence avoids the bias associated with it. VBA can also be automated with little human intervention.

Despite its advantages, VBA has some inherent limitations that affect its reliability. First, the effectiveness of VBA relies on the accurate registration of the images to a common space, which can be difficult to achieve in practice. The accuracy of registration is influenced by a number of factors, e.g, noise, artifacts, and complexity of brain structures (Hill et al., 2001). A perfect alignment is seldom possible in real-world applications. Registration errors may be taken as anatomical differences, inducing false positives in group comparison (Henley et al., 2010). Second, a sufficient amount of samples is needed to achieve reasonable statistical power. However, this requirement cannot always be met in view of low disease prevalence, recruitment difficulties, and demographic matching issues.

To overcome these limitations, we propose a method called block-based statistic (BBS) for robust group comparison. BBS leverages non-local means (NLM) (Buades et al., 2005) and non-parametric permutation test (Efron and Tibshirani, 1994) for robust voxel-wise statistical comparisons between two groups of images. We first perform the NLM-like block matching to determine matching structures, which we then leverage for group comparison. This mitigates the effects of comparing across disparate structures. We choose the non-parametric permutation test (Nichols and Holmes, 2002) for statistical inference since it makes little assumption about the distribution of a test statistic and therefore allows a non-parametric determination of group differences. The synthetic data experimental results demonstrate the effective of BBS quantitatively. Further real data experiments on the diffusion MRI data of mild cognitive impairment (MCI) patients show that BBS significantly improves statistical power.

### 1.1 Related Work

Our work is inspired by the non-local means (NLM) algorithm (Buades et al., 2005). Based on the observation that natural images contain plenty of recurrent structures, NLM removes noise by harnessing self-similar local image information. To identify self-similar information, block matching is performed, assigning a weight to each image block to indicate its similarity with reference to a target block. These weights are then used for estimating the noiseless signal of the voxel central to target block via weighted averaging. Block matching is also a powerful tool to correct for registration errors; see (Ourselin et al., 2000) for example. We will show in this work that block matching can be used to improve group comparison.

The fact that NLM can be used for improving power, robustness, and reproducibility of anomaly detection has been reported in (Commowick and Stamm, 2012). Compared with (Commowick and Stamm, 2012), our work has the following important distinctions:

1. *Group Comparison* – Our work deals with effective comparison between two groups. In contrast, (Commowick and Stamm, 2012) deals with comparison of an individual with a group.
2. *Non-Parametric Testing* – We do not rely on Gaussian assumption and use a non-parametric resampling-based approach for group comparison. In contrast, (Commowick and Stamm, 2012) inherently assumes Gaussian distribution by relying on the chi-squared distribution.
3. *Correction for Multiple Testing* – We incorporate a resampling-based procedure (Westfall and Young, 1993) to control for false-positive family-wise error rate (FWER) by taking into account the dependence structure between test statistics. Correction for multiple testing is not considered in (Commowick and Stamm, 2012).

A preliminary version of this work has been presented in a workshop (Chen et al., 2015). The method presented herein is significantly improved with more advanced block matching and correction for multiple testing. A more comprehensive evaluation of the proposed method is carried out with a sample size that is five times larger than the one used in our workshop paper. All these new materials are not part of our workshop publication.

## 1.2 Contributions

We introduce a robust block-based non-parametric statistical comparison method for improving VBA. Block matching is utilized both to correct for registration errors and to gather matching information for increasing sample size. Non-parametric permutation-based hypothesis testing is then applied for group comparison without having to assume Gaussian distribution. We enhance the algorithm by incorporating multiple-testing correction using the *step-down minP* procedure (Ge et al., 2003), by introducing a better means of computing the $p$-value, specific to the resampling-based procedure, using the concept of *exact Monte Carlo p-value* (Phipson and Smyth, 2010), and by utilizing a clustering-based method for unbiased selection of image references for block matching. Notable highlights of our method are summarized below:

1. *Sample Size Enhancement* – Statistical comparisons rely on sufficiently large sample sizes. Our method increases sample size by gathering similar information using block matching. Block matching is performed based on reference image groups determined using a clustering-based approach to better account for within-group heterogeneity.

2. *Correction for Registration Errors* – Registration errors introduce structural variability that will decrease the statistical power in detecting real meaningful differences. Our block-matching mechanism explicitly corrects for misalignment errors to ensure that comparisons are performed only between matching structures. Block matching also allows for soft correspondences between structures by assigning weights to structures based on their similarity to a reference structure. Since variability due to registration errors are minimized, our method significantly increases statistical power in detecting abnormalities.

3. *Block-Based Non-Parametric Permutation Statistics* – Since the Gaussian assumption does not necessarily hold in practice, we employ permutation test for group comparison. Unlike conventional permutation test, our method is done with consideration of the results given by block matching. We show that this will result in greater sensitivity to group differences while maintaining comparable specificity. To allow multivariate comparison, we choose the test statistic to be the Hotelling $T^2$ statistic (Anderson, 1958), which has been adopted in deformation-based morphometry (Gaser et al., 1999). To avoid underestimation of the $p$-value typical in permutation algorithms, we compute the exact $p$-value in the way described in (Phipson and Smyth, 2010).

4. *Resampling-Based Correction for Multiple Testing* – Consistent with the permutation testing framework described above, we utilize a resampled-based approach (Westfall and Young, 1993) for FWER control. This procedure, called step-down minP (Westfall and Young, 1993), takes a less conservative multi-step approach, in contrast to the single-step approach (Westfall and Young, 1993), to control for FWER based on the minimum $p$-value. Since the original procedure is computational expensive, we base our implementation on (Ge et al., 2003), producing the same outcome in a lesser amount of time. The step-down minP procedure does not rely on parametric assumptions on the joint distribution of test statistics under the null hypothesis.

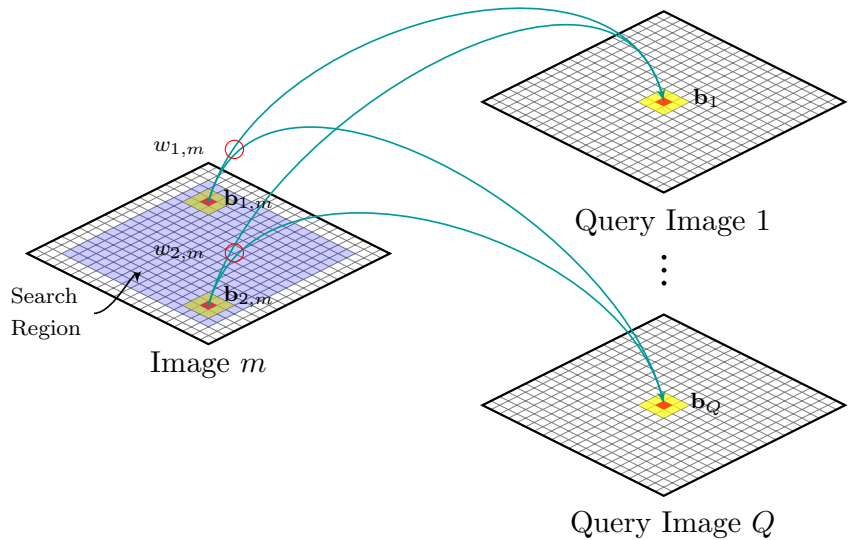

**Figure 1. Block Matching.** Block matching is performed based on a set of query images. Each block $\mathbf{b}_{i,m}$ within the search region in an image $m$ is compared with query block set $\{\mathbf{b}_q | q = 1, 2, \ldots, Q\}$. The weight $w_{i,m}$ indicates the degree of similarity between $\mathbf{b}_{i,m}$ and $\{\mathbf{b}_q\}$. Note that actual block matching is performed using 3D blocks.

### 1.3 Paper Organization

The structure of the paper is a follows. In Section 2, we describe the key components of the proposed approach. In Section 3, we demonstrate the effectiveness of the proposed algorithm with both synthetic and real data. Finally, in Section 4, we conclude this work.

## 2 METHOD

BBS consists of two major components: Block matching and permutation test. The first component corrects for alignment errors, and the second component utilizes matching blocks for effective non-parametric statistical inference.

### 2.1 Block Matching

We utilize block matching to determine similar structures for comparison purposes, reducing the influence of misalignments. Restricting statistical comparisons to only matched blocks will encourage comparison of similar and not mismatched information (e.g., due to structural misalignment). For group comparison, block matching is facilitated by a set of query images that are representative of the whole group. This is to avoid the bias involved in only using any single image from each group as the reference for block matching. When the group size is small, the query set consists of all images in the group. When the group size is large, a small number of query images representing different cluster centers can be selected with the help of a clustering algorithm. For any image in the group, block matching is performed with respect to the query images (see Fig. 1). That is, at each location in the common space of the query images, a set of blocks are concurrently compared with a block in the image.

We assume that two groups of images (i.e., $M_1$ images $I^{[1]}_{m_1 \in \{1, \ldots, M_1\}}$ in the first group and $M_2$ images $I^{[2]}_{m_2 \in \{1, \ldots, M_2\}}$ in the second group), which can be vector-valued, have been registered to the common space. Together, $\{I_{m_1}^{[1]}\}$ and $\{I_{m_2}^{[2]}\}$ constitute the complete datasets $I_{m \in \{1, \ldots, M\}}$ where $M = M_1 + M_2$. We are interested in comparing voxel-by-voxel images in the first group with images in the second group. For each point $\mathbf{x} \in \mathbb{R}^3$ in the common space defined by the query images of the two groups, we define a common block neighborhood $\mathcal{N}(\mathbf{x})$ and arrange the elements (e.g., intensity values) of the voxels in this block neighborhood lexicographically as two sets of vectors $\{\mathbf{b}_{q_1}^{[1]} \in \mathbb{R}^d | q_1 = 1, 2, \ldots, Q_1\}$ and $\{\mathbf{b}_{q_2}^{[2]} \in \mathbb{R}^d | q_2 = 1, 2, \ldots, Q_2\}$, where $Q_g$ is the number of query images in group $g \in \{1, 2\}$. We further define $\{\mathbf{b}_q\} = \{\mathbf{b}_{q_1}^{[1]}\} \cup \{\mathbf{b}_{q_2}^{[2]}\} = \{\mathbf{b}_q \in \mathbb{R}^d | q = 1, 2, \ldots, Q\}$ where $Q = Q_1 + Q_2$. Block matching is then performed as follows:

1. For each image $I_m$, search for blocks $\{\mathbf{b}_{i,m} | i = 1, 2, \ldots\}$ that are similar to block set $\{\mathbf{b}_q\}$.

2. Let $\mathbf{p}_{i,m}$ be a vector containing all the signals in the center voxel of $\mathbf{b}_{i,m}$, assign a weight $w_{i,m}$ to the central voxel $\mathbf{p}_{i,m}$ of each block $\mathbf{b}_{i,m}$, depending on the similarity between $\mathbf{b}_{i,m}$ and $\{\mathbf{b}_q\}$.

3. Based on the group that each sample belongs to, divide the weighted samples $\{(w_{i,m}, \mathbf{p}_{i,m}) | i = 1, 2, \ldots; m = 1, 2, \ldots, M\}$ into two groups $\{(w_{i_1, m_1}^{[1]}, \mathbf{p}_{i_1, m_1}^{[1]})\}$ and $\{(w_{i_2, m_2}^{[2]}, \mathbf{p}_{i_2, m_2}^{[2]})\}$, where

$$
\begin{aligned}
&\{(w_{i_1, m_1}^{[1]}, \mathbf{p}_{i_1, m_1}^{[1]}) | i_1 = 1, 2, \ldots; m_1 = 1, 2, \ldots, M_1\}, \\
&\{(w_{i_2, m_2}^{[2]}, \mathbf{p}_{i_2, m_2}^{[2]}) | i_2 = 1, 2, \ldots; m_2 = 1, 2, \ldots, M_2\},
\end{aligned}
\tag{1}
$$

4. Utilize two sets of weighted samples $\{(w_{i_1, m_1}^{[1]}, \mathbf{p}_{i_1, m_1}^{[1]})\}$ and $\{(w_{i_2, m_2}^{[2]}, \mathbf{p}_{i_2, m_2}^{[2]})\}$ to infer the differences between $\{I_{m_1}^{[1]} | m_1 = 1, 2, \ldots, M_1\}$ and $\{I_{m_2}^{[2]} | m_2 = 1, 2, \ldots, M_2\}$.

A straightforward approach to defining the weight is based on the distances between a candidate block $\mathbf{b}$ and the query block set $\{\mathbf{b}_q\}$. We define the weight based on the top $K$ best matching blocks $\{\mathbf{b}_{q_k}\}_{k=1}^K$ determined based on the Euclidean distance:

$$
w = \left[ \prod_{k=1}^K K_{\mathbf{H}} \left( \mathbf{b} - \mathbf{b}_{q_k} \right) \right]^{\frac{1}{K}},
\tag{2}
$$

where $K_{\mathbf{H}}(\cdot) = |\mathbf{H}|^{-1} K(\mathbf{H}^{-1} \cdot)$ is a multivariate kernel function with symmetric positive-definite bandwidth matrix $\mathbf{H}$ (Härdle and Müller, 2000; Yap et al., 2014). The weight indicates the similarity between a pair of blocks in a $(d + 3)$-dimensional space, where $d$ is the size of the image block $\mathbf{b}$, i.e., $d = |\mathbf{b}|$. To avoid small weights of mismatching blocks from influencing the results, only the top $L$ largest weights, denoted as $\{w_{i_g, m_g}^{[g]} | i_g = 1, 2, \ldots, L\}$, $g \in \{1, 2\}$, are used for permutation test.

Block matching helps correct for potential registration errors between images and increases the number of samples required for effective voxel-wise comparison. It also helps improve statistical power since confounding variability due to misalignment can be reduced.

## 2.2 Permutation Test

For the weighted samples $\mathbf{z}_{n_1}^{[1]} \equiv (w_{i_1, m_1}^{[1]}, \mathbf{p}_{i_1, m_1}^{[1]})$ and $\mathbf{z}_{n_2}^{[2]} \equiv (w_{i_2, m_2}^{[2]}, \mathbf{p}_{i_2, m_2}^{[2]})$ determined previously, we assume that they are independent random samples drawn from two possibly different probability distributions $F^{[1]}$ and $F^{[2]}$, i.e.,

$$
\begin{aligned}
F^{[1]} \to \mathbf{Z}^{[1]} &= (\mathbf{z}_1^{[1]}, \mathbf{z}_2^{[1]}, \dots, \mathbf{z}_{N_1}^{[1]}), \\
F^{[2]} \to \mathbf{Z}^{[2]} &= (\mathbf{z}_1^{[2]}, \mathbf{z}_2^{[2]}, \dots, \mathbf{z}_{N_2}^{[2]}).
\end{aligned}
\tag{3}
$$

Our goal is to test the null hypothesis $H_0$ of no difference between $F^{[1]}$ and $F^{[2]}$, i.e., $H_0 : F^{[1]} = F^{[2]}$. A hypothesis test is carried out to decide whether the data decisively reject $H_0$. This requires a test statistic $\hat{\theta} = s(\mathbf{Z}^{[1]}, \mathbf{Z}^{[2]})$, such as the difference of means. In this case, the larger value of the statistic, the stronger is the evidence against $H_0$. If the null hypothesis $H_0$ is not true, we expect to observe larger values of $\hat{\theta}$ than if $H_0$ is true. The hypothesis test of $H_0$ consists of computing the $p$-value of the test, and seeing if it is too small according to certain conventional thresholds. Having observed $\hat{\theta}$, the $p$-value is defined to be the probability of observing at least that large a value when the null hypothesis is true: $p$-value $= \text{Prob}_{H_0}\{\hat{\theta}^* \geq \hat{\theta}\}$. The smaller the value of $p$-value, the stronger the evidence against $H_0$.

The permutation test assumes that under null hypothesis $F^{[1]} = F^{[2]}$, the samples in both groups could have come equally well from either of the distributions. In other words, the labels of the samples are exchangeable. Therefore the null hypothesis distribution can be estimated by combining all the $N_1 + N_2$ samples from both groups into a pool and then repeating the following process for a large number of times $B$:

1. Take $N_1$ samples without replacement to form the first group, i.e., $\mathbf{Z}^{*[1]}$, and leave the remaining $N_2$ samples to form the second group, i.e., $\mathbf{Z}^{*[2]}$.

2. Compute a *permutation replication* of $\hat{\theta}$, i.e., $\hat{\theta}^* = s(\mathbf{Z}^{*[1]}, \mathbf{Z}^{*[2]})$.

We approximate the null hypothesis distribution by assigning equal probability on each permutation replication. The $p$-value is conventionally defined as the fraction of the number of $\hat{\theta}^*$ that exceeds $\hat{\theta}$, i.e., $p = \#\{\hat{\theta}^* \geq \hat{\theta}\}/B$. However, this definition can potentially lead to zero $p$-value, causing type I error that cannot be reduced by any multiple-testing correction methods. To overcome this problem, we compute instead the *exact $p$-value* as proposed in (Phipson and Smyth, 2010), i.e.,

$$
p = \frac{\#\{\hat{\theta}^* \geq \hat{\theta}\} + 1}{B + 1}.
\tag{4}
$$

## 2.3  Choice of Kernel

In general, there exist a variety of kernel functions for selection (Härdle, 1992). Consistent with non-local means (Buades et al., 2005), we use a Gaussian kernel, i.e., $K(\mathbf{u}) = \alpha \exp\left(-\frac{1}{2}\mathbf{u}^{\mathsf{T}}\mathbf{u}\right)$, and hence

$$
K_{\mathbf{H}}(\mathbf{u}) = |\mathbf{H}|^{-1} K(\mathbf{H}^{-1}\mathbf{u}) = \frac{\alpha}{|\mathbf{H}|} \exp\left(-\frac{1}{2}\mathbf{u}^{\mathsf{T}}\mathbf{H}^{-2}\mathbf{u}\right),
\tag{5}
$$

where $\alpha$ is a constant to ensure unit integral. The choice of $\mathbf{H}$ is dependent on the application. For simplicity, we set $\mathbf{H} = \text{diag}(h_1^{[\mathbf{b}]}, \dots, h_d^{[\mathbf{b}]}, h_1^{[\mathbf{x}]}, \dots, h_3^{[\mathbf{x}]})$ with $h_k^{[\mathbf{b}]} = \sigma\sqrt{d}$. The noise level $\sigma$ can be estimated by the method outlined in (Manjón et al., 2008). We set $h_k^{[\mathbf{x}]}$ to be half the value of the search radius.

## 2.4   Determining References for Block Matching

To take into consideration the heterogeneity that might occur within a group of images, we employ an image clustering approach to identify exemplar images that are representative of image clusters that are more homogeneous in the group. These exemplars are then used as references for block matching, which are referred to as query images in Section 2.1. This approach helps avoid the bias associated with selecting only one image as the reference.

A large number of clustering algorithms are available for this purpose (MacQueen et al., 1967; Bezdek et al., 1984; Frey and Dueck, 2007). In our work, we have chosen to use affinity propagation (AP) clustering (Frey and Dueck, 2007), which uses message propagation between data points to determine data clusters and their corresponding exemplars. Notably, AP clustering does not require the number of clusters to be pre-specified and determines it automatically based on the data.

## 2.5   Choice of Test Statistic

Diffusion MRI data can be represented as a vector-valued image with each voxel consisting of a vector with elements corresponding to a set of gradient direction and strength. To account for this vector-valued nature of the data, we use the multivariate Hotelling $T^2$ statistic (Anderson, 1958), which is known as a generalization of the Student's $t$ statistic. Leaving out the spatial location variable $\mathbf{x}$ for simplicity, the Hotelling $T^2$ statistic is defined as

$$T^2 = N(\bar{\mathbf{I}}^{[1]} - \bar{\mathbf{I}}^{[2]})^{\mathsf{T}} \mathbf{S}^{-1} (\bar{\mathbf{I}}^{[1]} - \bar{\mathbf{I}}^{[2]}), \tag{6}$$

where $\bar{\mathbf{I}}^{[g]}$ is the mean vector for group $g$, $\mathbf{S}$ is a covariance matrix, and $N$ is the number of samples. Based on the weighted samples obtained from block matching, the mean is computed via weighted averaging, i.e.,

$$\bar{\mathbf{I}}^{[g]} = \frac{\sum_{(w_{i,m}, \mathbf{p}_{i,m}) \in \mathbf{Z}^{[g]}} w_{i,m} \mathbf{p}_{i,m}}{\sum_{(w_{i,m}, \mathbf{p}_{i,m}) \in \mathbf{Z}^{[g]}} w_{i,m}}. \tag{7}$$

For a small number of samples, the empirical estimate of the covariance matrix $\mathbf{S}$ becomes unstable and singular, i.e., it cannot be inverted to compute the precision matrix $\mathbf{S}^{-1}$ (Schäfer and Strimmer, 2005). We approximate $\mathbf{S}$ by computing the variance matrix instead, denoted as $\hat{\mathbf{S}}$. In our case, we are interest in testing whether the differences between two groups are significantly greater than null. Therefore the variance matrix is computed as the summation of the variance matrices of the two groups, i.e., $\hat{\mathbf{S}} = \hat{\mathbf{S}}^{[1]} + \hat{\mathbf{S}}^{[2]}$. The variance matrix $\hat{\mathbf{S}}^{[g]}$ is a diagonal matrix. The diagonal elements of matrix $\hat{\mathbf{S}}^{[g]}$ is represented as a vector $\mathbf{s}^{[g]}$, which can be computed as

$$\mathbf{s}^{[g]} = \frac{\sum_{(w_{i,m}, \mathbf{p}_{i,m}) \in \mathbf{Z}^{[g]}} w_{i,m} (\mathbf{p}_{i,m} - \bar{\mathbf{I}}^{[g]}) \odot (\mathbf{p}_{i,m} - \bar{\mathbf{I}}^{[g]})}{\sum_{(w_{i,m}, \mathbf{p}_{i,m}) \in \mathbf{Z}^{[g]}} w_{i,m}}, \tag{8}$$

where $\odot$ denotes element-wise multiplication. A potentially more accurate estimation of the covariance matrix can be obtained using a shrinkage estimator (Schäfer and Strimmer, 2005).

Note that we choose to use the $T^2$ statistic because it is asymptotically pivotal (Hall and Wilson, 1991). Hypothesis testing using our permutation framework does not assume that the statistic is drawn from a Gaussian distribution.

---

**Algorithm 1** Step-down minP procedure

---

**Require:** Statistic matrix $\hat{\Theta}^*$ and raw $p$-value vector $\mathbf{p}$.
1: Compute $\mathbf{P}^*$ using the fast algorithm described in (Ge et al., 2003).
2: Order values in $\mathbf{p}$ such that $p_{s_1} \leq p_{s_2} \leq \ldots \leq p_{s_N}$.
3: Let $q^*_{N,b} = p^*_{s_N,b}$
4: **for** $n = N - 1$ **to** 1 **do**
5:    **for** $b = 1$ **to** $B$ **do**
6:       $q^*_{n,b} = \min\{q^*_{n+1,b}, p^*_{s_n,b}\}$
7:    **end for**
8:    $\widetilde{p}_{s_n} = \#\{p_{s_n} \leq q^*_{n,b}\}/B$
9: **end for**
10: Enforce the monotonicity constraint as follows:
11: **for** $n = 2$ **to** $N$ **do**
12:    $\widetilde{p}_{s_n} = \max\{\widetilde{p}_{s_{n-1}}, \widetilde{p}_{s_n}\}$
13: **end for**
14: **return** $\widetilde{\mathbf{p}}$

---

## 2.6 Correction for Multiple Testing

Since brain images typically consist of hundreds of thousands of voxels, counteracting the problem of multiple testing is a significant issue. Control for false positives in multiple testing include defining an appropriate false-positive error rate and devising multiple testing procedures that control this error rate by considering the joint distribution of the test statistics. Two popular error rate measures are the family-wise error rate (FWER) and the false discovery rate (FDR) (Efron, 2013).

Bonferroni correction is commonly used to control the FWER. It can however be somewhat conservative if there are a large number of tests and/or the test statistics are positively correlated. The correction also comes at the cost of increasing the probability of producing false negatives, and consequently reducing statistical power. Benjamini and Hochberg (Benjamini and Hochberg, 1995) proposed the FDR as an alternative to the FWER. With the FDR, more putatively significant hypotheses can be identified compared to the FWER. The authors proved in (Benjamini and Hochberg, 1995) that their procedure provides strong control of the FDR when the p-values from the true null hypotheses are independent.

In this work, we use a resampling-based method to control FWER (Westfall and Young, 1993). Our approach takes advantage of the permutation test to estimate the distribution of the test statistics under null hypothesis without relying on any prior distribution assumptions. The resampling-based approach can be based on either the maximum statistic (maxT) or the minimum $p$-value (minP). The maxT approach is generally faster than the minP method because the latter requires double permutation (Ge et al., 2003). However, this approach relies on the assumption that the test statistics are identically distributed across hypotheses and if this assumption is not satisfied, the different hypotheses might not be weighted equally (Ge et al., 2003). To avoid this, we implement the minP approach using the fast algorithm proposed in (Ge et al., 2003).

*Single-step* procedures, such as Bonferroni correction, adjust *all* $p$-values according to the minimum $p$-value distribution (Westfall and Young, 1993). However, it is possible to improve the power of these procedures by making the adjusted $p$-values uniformly smaller using a *multi-step*, or more specifically *step-down*, approach. In each step, only the minimum $p$-value is adjusted using the minimum $p$-value distribution. The remaining $p$-values are adjusted according to smaller and smaller sets of $p$-values. This has the effect of making the adjusted $p$-values smaller, thereby improving the power of the method, while maintaining the same error rate protection.

**Table 1.** Contingency Table.

|                  | Test Positive | Test Negative |
| ---------------- | ------------- | ------------- |
| True Difference  | TP            | FN            |
| No Difference    | FP            | TN            |

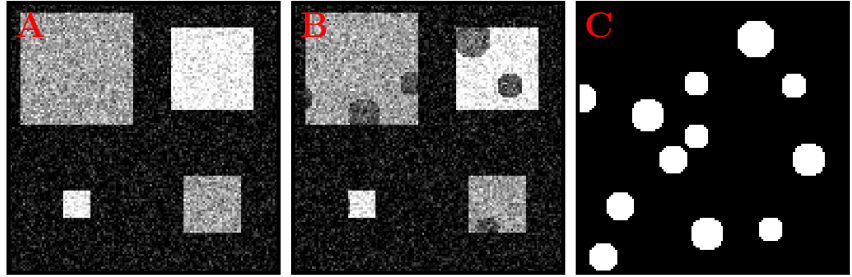

**Figure 2. Synthetic Data.** Left and Middle: Noisy reference datasets with normal structures (squares) and lesions (circles). Right: Regions with simulated lesions.

Suppose we have $N$ tests. After $B$ permutations, we obtain a matrix of test statistics

$$\hat{\mathbf{\Theta}}^* = \begin{bmatrix} \hat{\theta}^*_{1,1} & \hat{\theta}^*_{1,2} & \cdots & \hat{\theta}^*_{1,b} & \cdots & \hat{\theta}^*_{1,B} \\ \vdots & \vdots & & \vdots & & \vdots \\ \hat{\theta}^*_{n,1} & \hat{\theta}^*_{n,2} & \cdots & \hat{\theta}^*_{n,b} & \cdots & \hat{\theta}^*_{n,B} \\ \vdots & \vdots & & \vdots & & \vdots \\ \hat{\theta}^*_{N,1} & \hat{\theta}^*_{N,2} & \cdots & \hat{\theta}^*_{N,b} & \cdots & \hat{\theta}^*_{N,B} \end{bmatrix}. \tag{9}$$

Using the method described in (Ge et al., 2003), we can compute from $\hat{\mathbf{\Theta}}^*$ a matrix of raw $p$-values $\mathbf{P}^* = [p^*_{n,b}]$ and a matrix of minima of raw $p$-values $\mathbf{Q}^* = [q^*_{n,b}]$. Our objective is to compute the adjusted $p$-values, represented as vector $\widetilde{\mathbf{p}} = [\widetilde{p}_n]$, from the raw $p$-values $\mathbf{p} = [p_n]$ computed using (4). The *step-down minP* procedure is detailed in Algorithm 1.

## 3  EXPERIMENTAL RESULTS

We performed extensive experiments on synthetic and real *in vivo* diffusion MRI data to evaluate the effectiveness of BBS in detecting group differences using . The standard permutation test (Manly, 2006; Annis, 2005) was used as the comparison baseline. Instead of the weighted averaging scheme used in (7), for the standard permutation test, the mean is computed simply by averaging across images in the same group. Note that if we restrict the search range to $1 \times 1 \times 1$ and override the weights with 1, BBS is equivalent to the standard permutation test. For all experiments, we set the search range to $5 \times 5 \times 5$ and the block size to $3 \times 3 \times 3$. A search diameter of 5 is sufficient for correcting the registration errors because the images have already been non-linearly aligned. The number of permutations, $B$, was set to 2000. Voxels with corrected $p$-value below 0.01 were considered to be significantly different between the two groups.

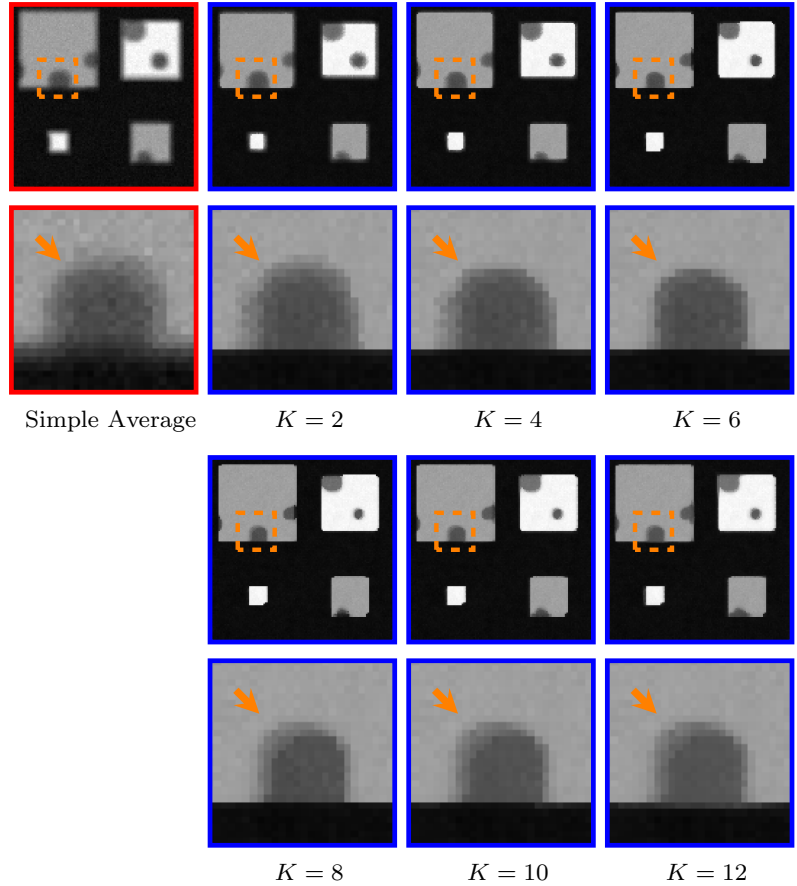

**Figure 3. Influence of** $K$**.** Left: Average image of 50 perturbed versions of reference patient dataset. The Rest: Average images after the block-matching correction with different $K$'s.

For quantitative evaluation, we used dice score, sensitivity, and specificity as measures. Based on Table 1, the dice score is defined as

$$\text{DS} = \frac{2|\text{TP}|}{2|\text{TP}| + |\text{FP}| + |\text{FN}|}, \tag{10}$$

sensitivity as

$$\text{SEN} = \frac{|\text{TP}|}{|\text{TP} + \text{FN}|}, \tag{11}$$

and specificity as

$$\text{SPC} = \frac{|\text{TN}|}{|\text{TN} + \text{FP}|}. \tag{12}$$

### 3.1 Synthetic Data

We first evaluate BBS using synthetic data. Based on a reference vector-valued diffusion-weighted image (each element corresponding to a diffusion gradient direction, see Fig. 2 (A)), 50 replicates were generated by varying the locations, sizes, and principal diffusion directions of the 'normal' structures (squares) to form the 'normal control' dataset. Based on a 'pathological' reference image (see Fig. 2 (B)), created by introducing lesions ('circles') to the reference image, a corresponding 'patient' dataset was generated by varying the locations, sizes, and severity of lesions, in addition to perturbing the normal structures as before. Lesions were simulated by swelling tensors in the radial directions.

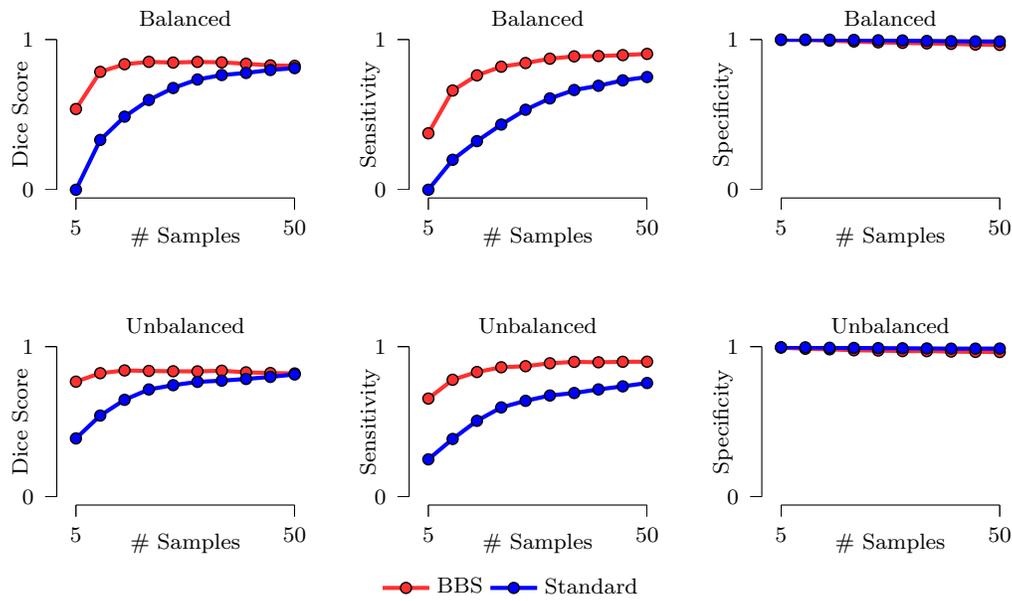

**Figure 4. Performance Statistics in Synthetic Dataset 1.** Detection accuracy, sensitivity, and specificity of BBS compared with the standard permutation test. The mean values of 10 repetitions are shown. The standard deviations are negligible and are not displayed. For the case of balanced sample size, both groups have the same number of samples. For the case of unbalanced sample size, only the size of the patient dataset was varied; the size of the normal control dataset was fixed at 50.

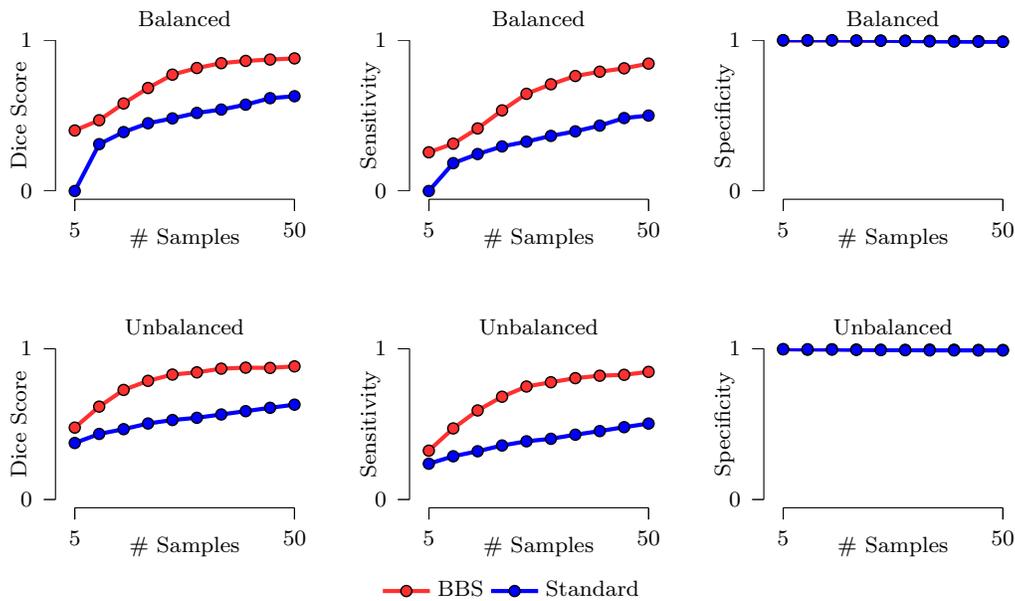

**Figure 5. Performance Statistics in Synthetic Dataset 2.** Similar to Fig. 4, but using Dataset 2.

We note that the performance of group comparison algorithms is influenced by the effect size. A conventional definition for effect size is the standardized mean difference between two groups, i.e., $\Phi = (\mu_1 - \mu_2)/\sigma$, where $\mu_1$ is the mean for one group, $\mu_2$ is the mean for the other group, and $\sigma$ is a standard deviation based on either or both groups. To evaluate the performance of our algorithm for different effect sizes, we changed $\sigma$ by adding different levels of noise to the ground truth images. Datasets 1, 2 and 3 were generated by respectively adding 6%, 8%, and 10% Rician noise. We simulated different levels of Rician noise by using the method proposed in (Coupé et al., 2008). Gaussian noise from distribution $\mathcal{N}(0, v(\theta/100))$ was added in the complex domain of the signal with percentage $\theta$. $v$ is the maximum signal value (100 in our case).

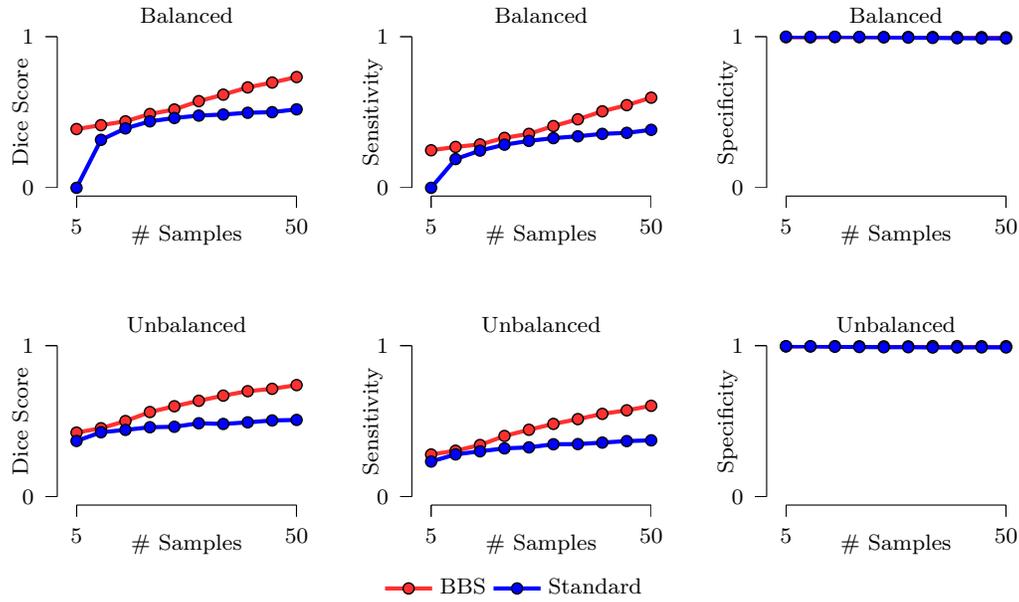

**Figure 6. Performance Statistics in Synthetic Dataset 3.** Similar to Fig. 4, but using Dataset 3.

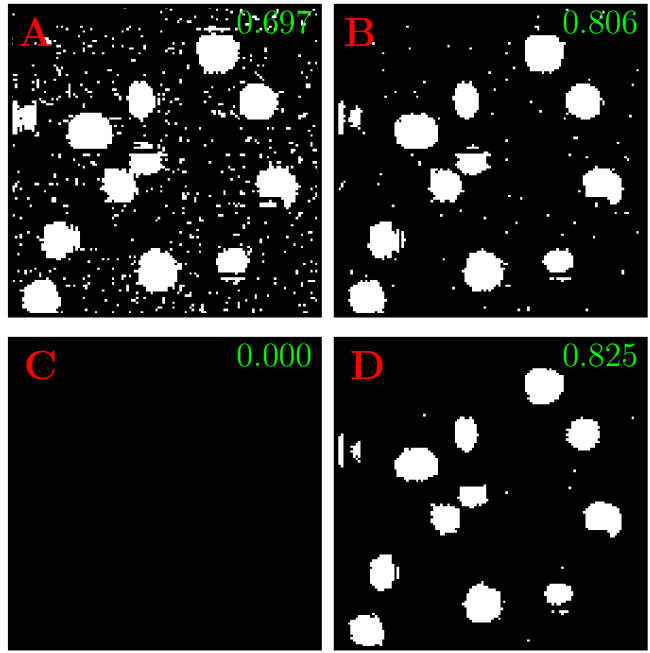

**Figure 7. Comparison of Correction Methods for Multiple Testing.** Detections given by BBS (A) without correction for multiple testing, (B) with FDR control, (C) with Bonferroni correction, and (D) with step-down minP correction. The value in the top right corner of each image is the dice score. The sample size is 40.

Before performing BBS, we first determined the cluster centers in each image group using AP clustering described in Section 2.4. The number of clusters is 6 in the normal control dataset and 7 in the patient dataset, resulting in a total number of $Q = 6 + 7 = 13$ reference. It can be observed from Fig. 3 that the best edge-preserving effect is obtained when $K = 6$, i.e., when $K = \min\{Q_g | g = 1, 2\}$. Therefore, we set $K = 6$ in the following synthetic data experiments.

To demonstrate the efficacy of BBS, we performed group comparison by progressively increasing the number of samples. Detection accuracy was evaluated using Dice score with the lesions defined on the reference image as the baseline (see Fig. 2 (C)). Both cases of balanced and unbalanced sample sizes were considered. For the latter, only the size of the patient dataset was varied; the size of the normal control dataset was fixed at 50. The results, shown in Fig. 4, 5 and 6, indicates that BBS yields markedly improved accuracy even when the sample size is small. Detection sensitivity is greatly increased by BBS. The specificity of both methods is comparable. The improvements given by BBS can be partly attributed to the fact that BBS explicitly corrects for alignment error.

Comparing the results in Fig. 4, 5 and 6, we can observe that, with the decrease in effect size (i.e., increase in noise level), the performance of the standard permutation test drops dramatically, while BBS gives consistent performance for all effect sizes.

We also compared the effectiveness of the step-down minP procedure with more conventional approaches, such as the Benjamini-Hochberg (BH) procedure for FDR control and Bonferroni correction. The results, shown in Fig. 7, indicate that the step-down minP procedure gives the best control over false positives while at the same time retain high detection rate. The BH procedure still retains a significant amount of false positives. Bonferroni correction is too conservative and no difference is deemed significant. The superiority of the step-down minP procedure is confirmed by the dice scores shown at the top right corners – the step-down minP procedure gives the highest dice score.

Figure 7 also indicates that BBS manages to control the occurrence of false positives in the background where no group differences are expected to exist. This demonstrates that BBS is able to enhance statistical power without increasing the false-positive rate.

## 3.2 Real Data

The real data consist of 100 scans for 50 healthy subjects and 50 mild cognitive impairment (MCI) subjects recruited in Geneva and Lausanne counties, Switzerland. Informed written consent was obtained from the subjects, and the data acquisition was approved by the ethical committee of the University Hopsitals of Geneva, Switzerland. Diffusion MRI was performed on a clinical routine whole body Siemens 3T Tim Trio MR scanner with a standard sequence: 30 diffusion directions uniformly distributed on a hemisphere with $b = 1{,}000 \, \text{s/mm}^2$, one image with no diffusion weighting, $128 \times 128$ matrix, voxel size $2 \times 2 \times 2 \, \text{mm}^3$, TE $= 81 \, \text{ms}$, TR=7,618 ms, 1 average.

From the diffusion-weighted (DW) images of each subject, the fractional anisotropy (FA) and mean diffusivity (MD) were generated by fitting a diffusion tensor to each voxel. Non-linear fields deforming the subject native spaces to the FSL FA standard space[1] were estimated using diffeomorphic demons (Vercauteren et al., 2009) based on the FA images. Based on the estimated deformation fields, the DW, FA, and MD images were warped to the standard space (Yap and Shen, 2012).

---

[1] `http://fsl.fmrib.ox.ac.uk/fsl/fslwiki/FMRIB58_FA`

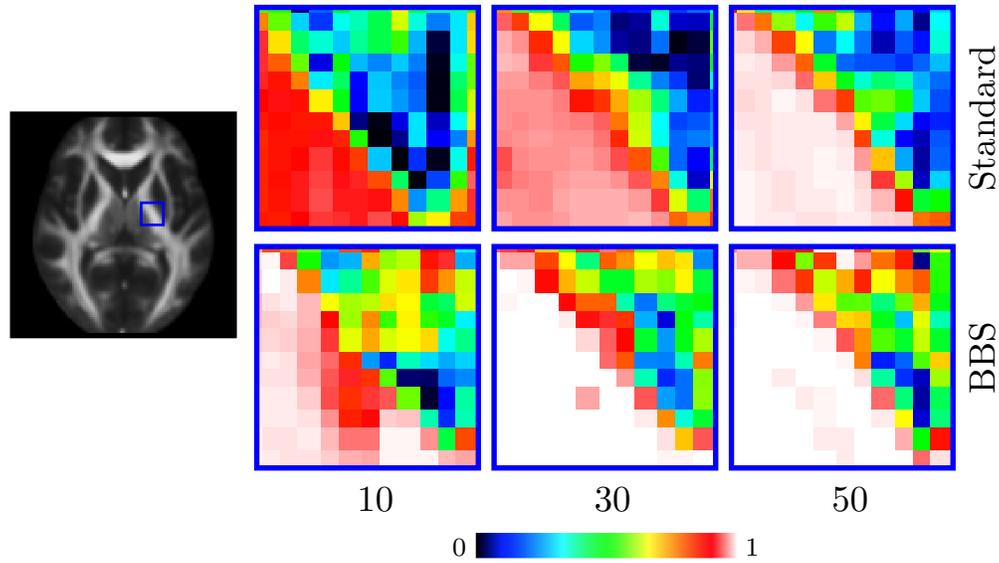

**Figure 8. Comparison of Significance Maps.** In each view, the FA image (far left) is shown for reference. The top row is given by the standard permutation test. The bottom row is given by BBS. From left to right, the sample size is 10, 30, and 50, respectively.

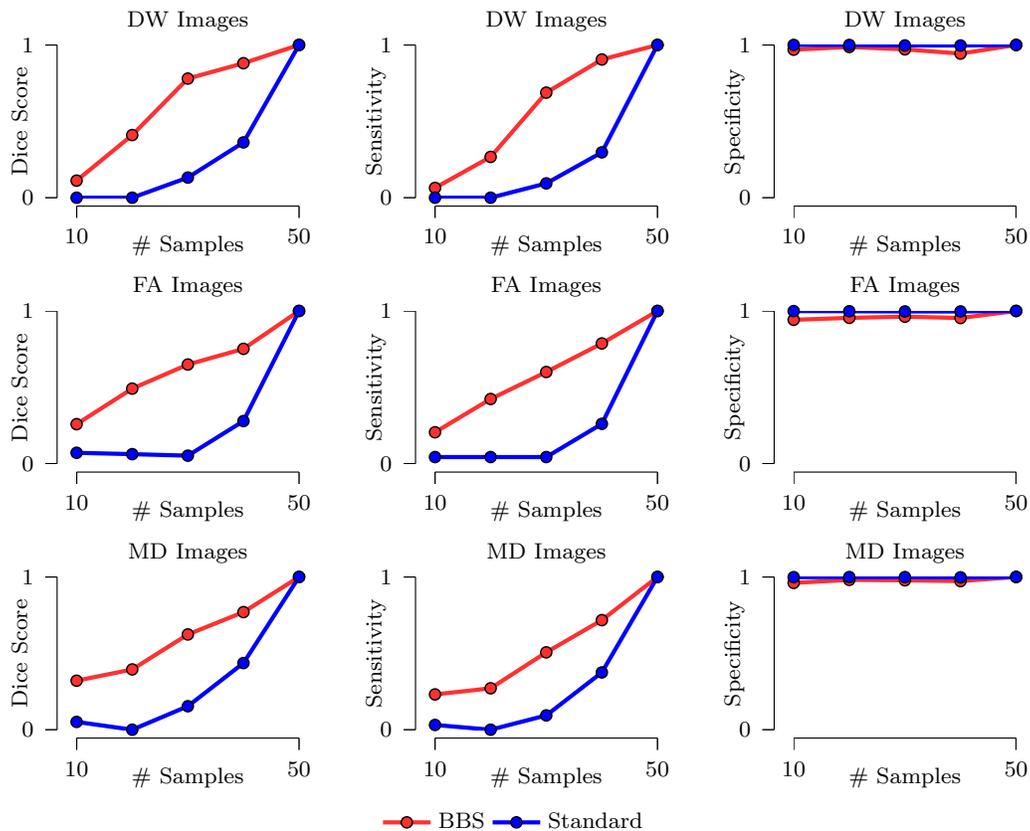

**Figure 9. Performance Statistics for Real Data.** Detection accuracy, sensitivity, and specificity of BBS compared with the standard permutation test. DW, FA, and MD images are used in the experiments. The mean values of 10 repetitions are shown. The standard deviations are negligible and are not displayed.

Experiments were carried out using the DW, FA, and MD images. We set $K = \min\{Q_g | g = 1, 2\} = 6$ based on AP clustering. We compared the performance of BBS and standard permutation test qualitatively and quantitatively. Representative visual comparison results based on DW images are shown in Fig. 8. The color images present the $(1 - p)$-values obtained by the two methods using 10, 30, and 50 images in each group. Warmer and brighter colors indicate differences with greater significance. When a limited number of samples are available, BBS gives results that are more consistent with those obtained using a larger sample size. We also note that the BBS results obtained with 10 samples are close to the standard permutation results obtained with 50 samples, implying that BBS is capable of increasing the sensitivity when only a small sample size is available.

For quantitative evaluation, we used the detection results obtained using the full 50 samples as the reference and evaluated whether hypothesis testing using a smaller number of samples gives consistent results. The evaluation results based on different image modalities, shown in Fig. 9, indicate that BBS yields results that converge faster to the results given by a larger sample size. This confirms that BBS improves group comparison sensitivity in small datasets.

## 4 CONCLUSION

We have proposed a new method, called block-based statistics (BBS), for detecting group differences with greater robustness. BBS leverages block matching to correct for image local misalignments. Based on the block matching outcome, permutation test is employed for robust non-parametric statistical inference. The effectiveness of BBS has been confirmed by extensive experiments on synthetic data and the real diffusion MRI data of MCI patients.

## CONFLICT OF INTEREST STATEMENT

The authors declare that the research was conducted in the absence of any commercial or financial relationships that could be construed as a potential conflict of interest.

## AUTHOR CONTRIBUTIONS

GC and P-TY implemented the code and designed the experiments. GC and P-TY drafted the manuscript. P-TY revised the manuscript. C-YW provided the clinical data. PZ, KL, WP, YW, and DS participated in idea discussion and reviewed the manuscript.

## FUNDING

This work was supported in part by a UNC start-up fund and NIH grants (NS093842, EB022880, EB006733, EB008374, EB009634, MH088520, AG041721, and MH100217). The first author was supported by a scholarship from the China Scholarship Council.